%% file: Kronoscope.tex
\documentclass[11pt,a4paper,reqno,nofootinbib]{article}
\input{preambulesobre}
\input{commandes}\headheight 1.2cm\textheight 24cm \textwidth 17cm 
\begin{document}
\title{\vspace{-.50truecm}Emergence of time in quantum gravity: \\
  \vspace{.15truecm} is time necessarily flowing ?}
 \author{Pierre Martinetti\\ {\small CMTP \& Dipartimento di Matematica, ~Universit\`a di Roma Tor
   Vergata, I-00133;}\\ {\small Dipartimento di fisica,~Universit\`a di Roma
   ``Sapienza'', I-00185.}\\\vspace{.25truecm}
\small{martinet@mat.uniroma2.it}}
\date{\emph{Contribution to the Workshop ``Temps et Emergence'', \\Ecole
    Normale Supérieure, Paris 14-15 october 2011. \\To be published in Kronoscope.}}
\maketitle

\abstract{
We discuss the emergence of time in quantum gravity, and ask whether time is always ``something that flows''. We
first recall that this is indeed the case in both relativity and quantum mechanics, although in very different manners: time flows
geometrically in relativity (i.e. as a flow of proper time in the
four dimensional space-time), time flows abstractly in quantum
mechanics (i.e. as a flow in the space of observables of the system).
We then ask the same question in quantum gravity, in the light of the
thermal time hypothesis of Connes and Rovelli. The latter proposes to
answer the \textit{question of time in quantum gravity }(or at least
one of its many aspects), by postulating that time is a state
\textcolor{black}{dependent} notion. This means that one is able to
make a notion of time-as-an-abstract-flow - that we call the
\textit{thermal time} - emerge from the knowledge of both:

\begin{enumerate}
\item [-]  the algebra of observables of the physical system under
  investigation,

\item [-]  a state of thermal equilibrium of this system.
\end{enumerate}
Formally, this thermal time is similar to the abstract flow of
time in quantum mechanics, but we show in various examples that it may
have a concrete implementation either as a geometrical flow, or as a
geometrical flow combined with a non-geometric action. This indicates
that in quantum gravity, time may well still be
``something that flows'' at some abstract
algebraic level, but this does not necessarily imply that time is
always and only ``something that flows'' at
the geometric level.}

{\footnotetext{The author is supported by the
ERC advanced grant 227458 OACFT \textit{Operator Algebra \& Conformal
Field Theory} and the ERG{}-Marie Curie fellowship 237927
\textit{Noncommutative geometry \& quantum gravity}.}}

\bigskip
\section{Introduction}

We question the notion of time{}-flow, and the way it emerges is some
approaches to quantum gravity. In particular, we ask whether time is
necessarily ``something that
flows''.

In a first part, we recall how, in relativity as well as in quantum mechanics, time is
indeed ``something that flows'', although the flows do not take place at the same levels: in relativity time is a
geometrical flow in space-time whereas, in quantum mechanics, it is a flow in an abstract
space, describing the algebraic structure of the theory. In quantum gravity, as shown in
the second part, these two figures of temporality need to be reconciled. A way of reconciliation is offered by thermodynamics. This
is the \textit{thermal time hypothesis} of Connes and Rovelli \cite{Connes:1994xy}, who propose to extract {\textit{a
    time-as-geometrical{}-flow}} from a
\textit{time-as-abstract-flow.} 
Mathematics and theoretical physics provide
a natural framework to test this idea. In the last part of this
contribution, we thus investigate various physical situations in which
the pertinence of the thermal time hypothesis is verified. In the first
two examples, the so-called Unruh
effect{\footnote{\ \textcolor{black}{The
Unruh effect is defined later on the text. In short, it is the
prediction that an accelerating observer holding a ``quantum
thermometer'' that measures the temperature of the quantum vacuum would
measure a non{}-zero temperature where an inertial observer would see a
zero temperature.}}
and a variation on the Unruh effect for
non-eternal observers, we find that the concrete physical
implementation of the thermal time is a physically meaningful
geometrical flow. We then exhibit a third example where the concrete realisation of the thermal time combines
the previous geometric flow on space-time
with a purely non-geometric action on the physical
observables. In
other terms, time is only partially flowing.

As a conclusion, we show that this
multiplicity of modes of emergence of time suggests non-trivial
answers to classical philosophical questions such as ``is
there something rather than nothing~?'', or to seemingly naive questions such as
``is there more light at day or
night ?''.

\section{Time-as-geometrical-flow vs. time-as-abstract-flow}

\subsection{The  geometric  time of  relativity}

In relativity, time-evolution has a clear
geometrical interpretation: the movement of an observer, that is the evolution of its
position as time passes, is described by a \emph{line of
universe}, namely a one-dimensional trajectory in a
mathematical space-time of dimension four. The distinction between
space and time, however, is not univocal. Each observer makes the separation of the
four-dimensional space-time into a three-dimensional space and a
one-dimensional time in his own way, depending on his state of
movement. For instance in Minkowski space, that is the flat space-time
of special relativity in the absence of
gravity, the temporal evolution of a static
observer is described by a line parallel to
the $T$ axis, and the surfaces of simultaneity (in a two dimensional
space-time, see figure \ref{kronos1}) are parallel to the $X$ axis. Meanwhile, for
an observer whose speed is constant with respect to the static observer, the temporal evolution as well as the surfaces of
simultaneity are no longer parallel to any of the axes. For an observer with constant acceleration, the line of universe is
an hyperboloid, and simultaneity surfaces are lines through the origin (see figure \ref{kronos2}).

\begin{figure}[h*]
\begin{center}
\vspace{0truecm}
\mbox{\rotatebox{0}{\scalebox{.4}{\hspace{1truecm}\includegraphics{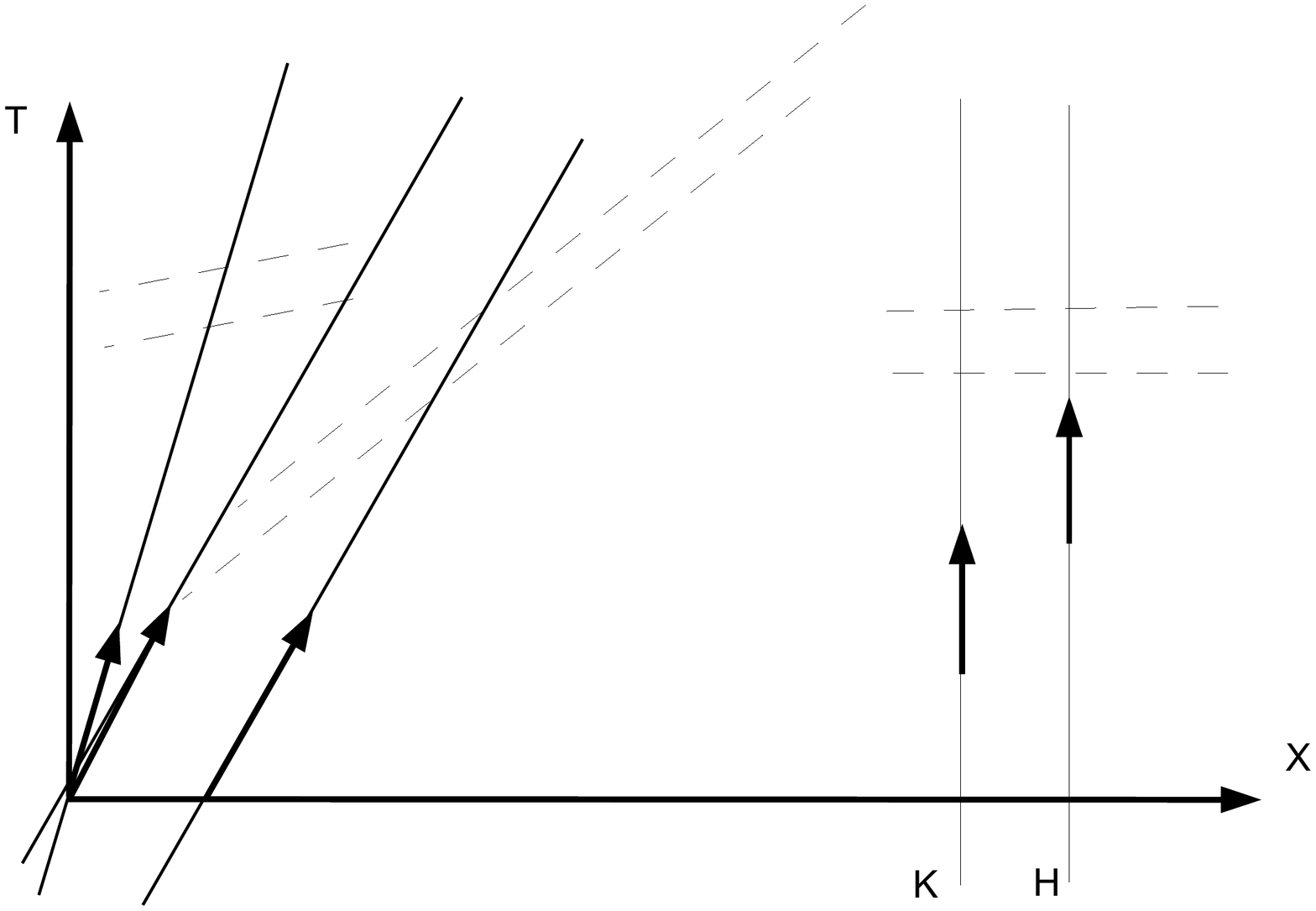}}}}
\vspace{-0.5truecm}
\begin{minipage}{.85\linewidth}{\caption{\label{kronos1}\emph{Separation of space and time in the 2-dimensional
Minkowski space-time}: the $T$ axis is the line of universe of an observer who stays at the same
place $X = 0$ at any time. In this space-time, being immobile at a
point $K$ or $H$ in space corresponds to lines of universe parallel to this
$T$ axis (on the right). On the left, the lines of universe of two
observers with the same non -zero constant speed (parallel lines), and
a third one with a lower speed. The dashed lines are the surfaces of
simultaneity. The latter are parallel to the $X$ axis only for static
observers. The arrows are the vectors tangent to the lines of universe.
Each of them has length $1$: the apparent difference between the
length of the arrows is an effect of hyperbolic geometry (Minkowski
space-time is not an Euclidean geometry).}}\end{minipage}
\end{center}
\end{figure}

 In other words, and unlike Newton mechanics, in relativity there is no
absolute time, that is to say no global object which flows
``everywhere in the same way''. Nevertheless,
although the relativistic time is not everywhere the same, it is
\emph{locally unique}. Each observer along
his line of universe does experiment a single time, his \textit{proper
time. }The proper time of a first observer may not be identical to the
proper time of a second observer, but one knows (in principle at least,
the explicit computations can be involved) how to go from one to the
other.

\begin{figure}[h*]
\begin{center}
\vspace{0truecm}
\mbox{\rotatebox{0}{\scalebox{.35}{\hspace{1truecm}\includegraphics{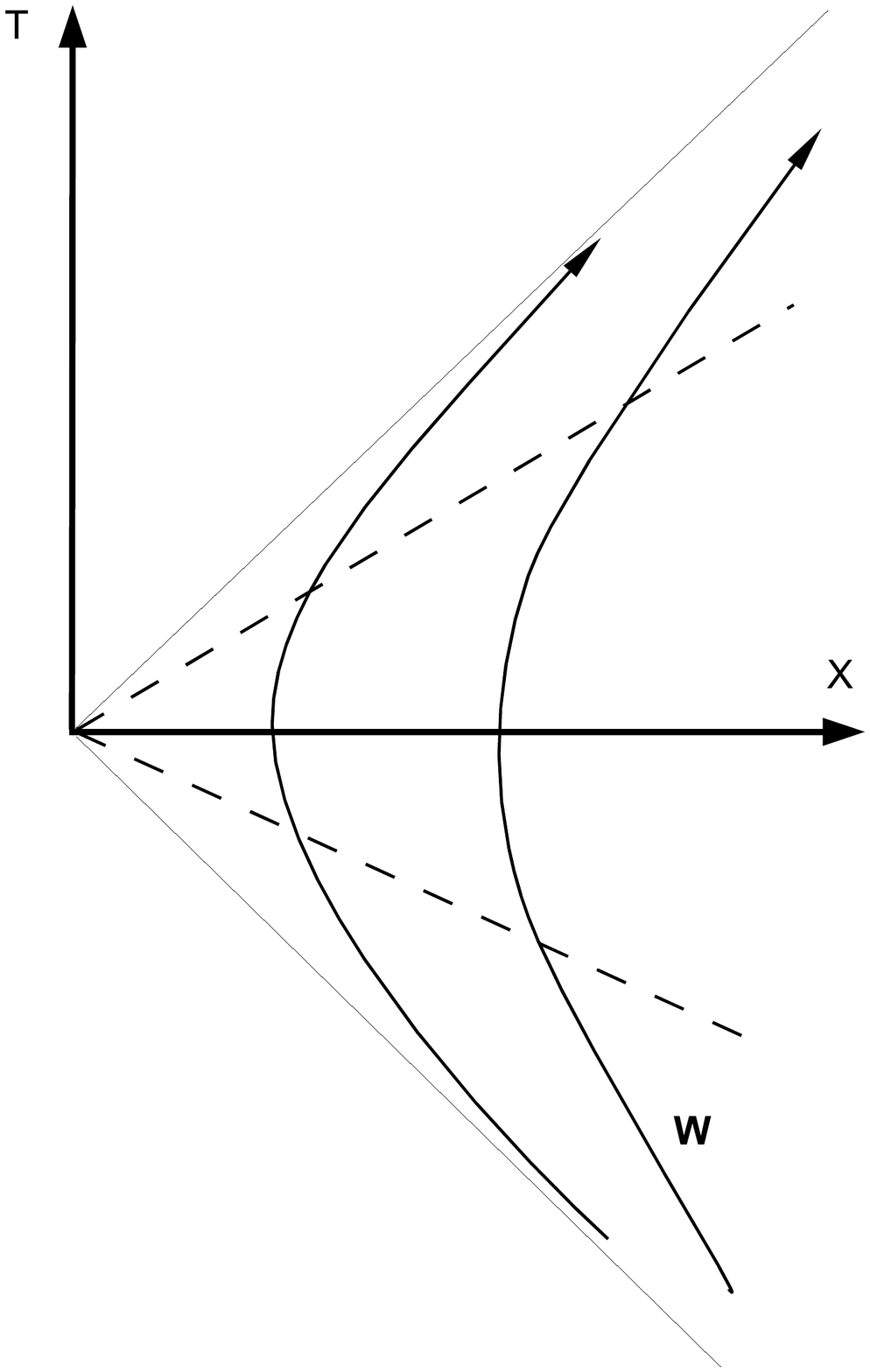}}}}
\vspace{-.5truecm}
\begin{minipage}{.75\linewidth}{\caption{\label{kronos2}\emph{The Rindler wedge}: the hyperboloids are the
lines of universe of observers with constant acceleration, and the
dashed lines are the surfaces of simultaneity. The two asymptotes to
these trajectories are lines of universe of light. The wedge W
encompassed by these two light-like trajectories is the region of
space-time causally connected to the accelerated observers.}}\end{minipage}
\end{center}
\end{figure}

 Geometrically, the unicity of the proper time means that along each
line of universe, at any point one has one and only one unit tangent
vector (here ``unit'' means of length 1).{\footnote{To avoid confusion, let us insist that the
notion of time flow should not be understood as \ a property of a
point, but as a property of a trajectory in space-time.: in figure 1,
at any \ given point one can attach as many arrows as lines of
universe that can go through this point: that is infinitely many ! }} 
In the absence of gravity, the switching of proper time between two
observers of constant speed is done thanks to some geometrical
transformations called \emph{Poincar\'e
transformations}. Consider for instance the lines of
universe of two static observers (the two lines on the right of figure \ref{kronos1}). A vector tangent to the first is mapped to a vector tangent to the second by the simplest
Poincar\'e transformation, that is a translation. The same is true for
two observers with the same constant speed (parallel lines on the left
of figure 1). For observers with different  - but constant  - speeds
(the two lines that intersect at the origin in figure 1), one uses
another Poincar\'e transformation to map a vector tangent to the first
curve to a vector tangent to the second one, namely a \emph{Lorentz
  boost} (in essence: a rotation in the T,X plane).

 To get an intuitive idea of what is going on, it could be useful to
draw a simple analogy with the usual time measured by our clocks and watches on Earth. There is no absolute time
on Earth: noon -bells do not ring simultaneously in Kathmandu or Haiti. But there is a universal time, divided in time-zones, and each
observer knows how to regulate his clock according to it: noon-bells should ring when the Sun is at the highest in the sky of Kathmandu or
Haiti. A traveller knows very well how to pass from one time-zone to another, by simply put one's watch forward or back a
few hours. Putting forward or back are precisely translations. In the geometrical picture above, this means that the origin of time on the
line of universe X= Kathmandu is not on the same surface of simultaneity as the origin of time on the line of universe X= Haiti.

To illustrate why the translations do not cover all the possible cases
and why other Poincar\'e transformations are required, let us imagine
that on Earth one takes into account a second system of time-zones,
for instance based on the Moon rather than the Sun. Inside this lunar
time system, the lunar noon is defined as the moment of time where the
Moon is highest in the sky. Each observer knows which lunar time-zone
he is belonging to, and he knows the lunar-time difference between
the lunar noon of Kathmandu and the lunar noon of Haiti. The change of
lunar time-zone is not more difficult than the usual change of solar
time-zone (in the same way as the mapping of tangent vectors between
two same-speed lines of universe is a translation, whatever the
speed). The difficulty arises when one intends to switch from a lunar
time-zone to a solar time-zone (in the same way: a translation does
not allow to map tangent vectors between two constant-speed lines of
universe with different speed). An observer in the solar time who aims at
regulating his clock on the lunar time can not simply put back or
forward his clock. He needs to modify all the inside mechanism of his
clock, for a lunar day cannot be divided into twenty four solar hours.
Nevertheless this change of time system is possible, and well described
by the formalism of relativity.

 When gravity enters the game and curves the space-time, the
multiplicity of time systems according to which one could regulate
one's clock might become even more involved
(technically speaking, this means there may exist many
\emph{time-like Killing vector fields}). On Earth, this is as if one were considering not only the solar or lunar times, but
imagine more systems, for instance based on Mars, Jupiter, Halley's comet etc. Each system can equally be
promoted as ''the'' time system, and the effective choice will be made on practical reasons: it is easy to
indicate solar time thanks to a sundial, it not easy to build a Jupiter-dial ! The same is true in relativity: on a given spacetime,
one will choose the coordinate system most appropriate to the system under study. But in any case, besides the diversity of the clocks used
by different observers to measure their own proper time, all observers
will agree: there is ``something flowing''.
Only the ways of flowing are different.

\subsection{The algebraic time of quantum mechanics}

In quantum mechanics, time is not a geometrical flow. Time-evolution
is charac\-terized as a transformation that preserves the algebraic
relations between physical observables. If at a time $t=
0$ an observable  - say the angular momentum
$L(0)$ - is defined as a certain combination (product
and sum) of some other observables  - for instance positions
$X(0)$, $Y(0)$ and momenta $P_X(0), P_Y(0)$, that is to say
\begin{equation}
L(0)= X(0) P_Y(0) - Y(0) P_X(0),
\end{equation}
then one asks that the same relation be satisfied at any other instant
\textit{t }(preceding or following $t=0$),
\begin{equation}
L(t)= X(t) P_Y(t) - Y(t) P_X(t).
\end{equation}

The quantum time-evolution is thus a map from an observable
at time $0$ to an observ\-able at time $t$ that
preserves the algebraic form of the relation between observables.
Technically speaking, one talks of an \textit{automorphism }of the
algebra of observables.

At first sight, this time-evolution has nothing to do with a flow.
However there is still ``something flowing'',
although in an abstract mathematical space. Indeed, to any value of $t$
(here time is an absolute parameter, as in Newton mechanics)
is associated an automorphism
$\alpha_t$ that allows to deduce the
observables at time $t$ from the
knowledge of the observables at time $0$. Mathematically, one writes
\begin{equation}
L(t) =\alpha_t(L(0)), X(t) = \alpha_t(X(0))
\end{equation}
and so on for the other observables.

The term ``group'' is important for it
precisely explains why it still makes sense to talk about a flow. Group
refers to the property of additivity of the evolution: going from
$t$ to $t'$ is equivalent to going from $t$
to $t_1$, then from $t_1$ to $t'$.
Considering small variations of time $\frac{t'- t}n$ where $n$ is an integer, in the
limit of large $n$ one finds that going from $t$ to
$t'$ consists in flowing through
$n$ small variations, exactly as the geometric flow consists in
going from a point $x$ to a point $y$ through a great
number of infinitesimal variations $\frac{x-y}n$.That is why
the time-evolution in quantum mechanics can be seen as a ``flow'' in
the (abstract) space of automorphisms of the algebra of observables.

To summarize, in quantum mechanics time is still
``something that flows'', although in a less
intuitive manner than in relativity. The idea of ``flow of
time'' makes sense, as a flow in an abstract space rather
than a geometrical flow.

\section{The question of time in quantum gravity}

As we have seen in the previous section, the time in relativity does not
flow in the same manner as the time in quantum mechanics: the former is
a geometrical flow in space-time, the latter is a 1 -parameter group
in the space of automorphisms of the quantum
observables. Combining
quantum physics with gravity thus inevitably yields to a conflict
between these two ``ways of flowing''.

The \textit{question of time }in quantum gravity takes many aspects
(see \cite{Saint-Ours:2011fk}, \cite{Saint-Ours:2011uq}). Here, we will consider the problem of the
unicity of the classical limit of the flow of time, when one passes
from quantum gravity to general relativity. By classical, we intend the
relativistic, non-quantum, limit of quantum gravity. In a rather
unexpected way, the abstract nature of the flow of time in quantum
mechanics  - rather than being a problem  - brings a solution.

\subsection{Uniqueness of the classical limit}

A founding principle of quantum mechanics may be summarized as
``everything is possible, but with a certain
probability''. The evolution of this probability is
per\-fectly deterministic. Quantum uncertainties appear when one
effectively measures a physical observable: the theory predicts the
probability that such or such measure gives such or such result, but it
does not predict the outcome of one single measure. Before the measurement
process, a quantum system is thus described by a\textit{ state }which
is the sum of all the possibles, each of it with a given probability.
We speak of \textit{a statistical state. }The most famous example is
the Schr\"odinger cat which, as long as no observer verifies his
health, is described as a superposition of a dead cat and a living cat
(this is called the principle of \textit{superposition of the wave
packet}).

In the same way, in quantum gravity one
expects the gravitational field to be described by a superposition of
all possibles. Since in general relativity the
gravita\-tional field determines the metric of space-time, and the
latter indicates how space can be separated from time, on expects that
at the quantum level any direction can be picked out as the
direction of time. But in general relativity, time is locally unique (as
explained in section II.1, an observer experiences one and only one
proper time). How can one reconcile the freedom given by quantum physics in the choice
of the time direction, with the unicity of the proper time, locally imposed by general relativity?

Coming back to the analogy with time on Earth, this is a bit as if, assuming the gravitational field in the Solar
System is completely determined by the mass of the Sun, one considers a quantum state of
the Solar system as a superposition of all the ``would be'' Solar Systems, corresponding to Suns with different
masses.{\footnote{Let
us warn the reader that this analogy, for what it be worth, only aims at giving a reader with no
particular background in physics a ``taste'' of what is going on. By no mean it should be taken as a serious
physical explanation. For instance we are mixing without any caution macroscopic and quantum effects, and from an astrophysical point of
view it does not make much sense to imagine a quantum superposition of
Suns with various masses!}}
Then,  a quantum-gravity clock should be described as a superposition of as
many clocks as possible values of the Sun mass. Each of these clocks
measures the ``something flowing''  discussed at the end
of section II.1 (whose concrete realization as a
proper time will depend on the observer), corresponding to a
value of the Sun mass. Locally, for an observer who has chosen once for all his time system  - say the solar
time  - the quantum-gravity clock will take the form of a quantum
sundial, that is a superposition of many sundials, each of them
measuring the solar time corresponding to one of the possible values of the Sun
mass. At the non-quantum limit, how does this quantum sundial collapse to the usual sundial?

Before examining a possible solution to this
problem, let us emphasize that the problematic is, in a sense, opposite to the one that might
have been expected at first sight: time in quantum mechanics is an
absolute parameter, unique, whereas proper-times in general relativity are
as many as observers. The principle of superposition of the wave packet, applied to the gravitational field,
reverses the problem: quantum mechanics offers the multiplicity, while
relativity, restricted to one single observer, asks for unicity.

\subsection{The thermal time hypothesis}

A solution to the uniqueness problem of the classical limit of the time
flow has been proposed by Connes and Rovelli under the name of
\textit{thermal time hypothesis} \cite{Connes:1994xy} (see also \cite{Rovelli:2010fk} for a recent analysis). The idea is to extract from an
equilibrium state the \textit{time-as-abstract-flow }of quantum
mechanics (i.e. a flow of automorphisms), then to turn it into the
locally unique \textit{time-as-geometrical-flow } of relativity.
The almost miraculous aspect of this proposition is that it has an
immediate translation in a branch of mathematics which has a priori
nothing to do with the question, namely the modular group in the theory
of von Neumann algebra. Quoting R. Haag \cite{Haag1996}, this
is an example of the remarkable'' pre-harmony between
mathematics and physics''.

To understand well the solution proposed by the thermal
time hypothesis, it could be useful to make a short digression
regarding the notion of state and equilibrium.
Let us begin by emphasizing the difference
between a physical system (e.g. the cat), and its states (e.g. a living
cat, or a dead cat, or a superposition of both). In a similar way as,
for Heidegger \cite{Heidegger:1962fk} it not so much the greek temple that is made of
stone than the stone that comes to existence through the temple, a
physical system comes to existence - in our case: become accessible
to the experiment - through its various possible states. The Schr\"odinger cat by itself has no existence, what appears in the equations of quantum
mechanics are the mathematical objects encoding its various possible states of existence (alive, dead, half alive/half dead). Describing a
system by a state is typical of thermodynamics. Consider a gas in a
box: one has no access to the movement of each atom of gas, however one
is able to obtain relevant physical information by studying the statistical behaviour of these atoms. Said differently, one does not
know the microscopic behavior of each particle (when will this precise particle hit the edge of the box ?),
however it makes sense to define, at the macroscopic scale, the pressure of the gas on the edge of the box. Thus, there is a
distinction between a macroscopic state - intended as a set of values of the thermodynamical quantities describing the system, like
 energy, temperature, pressure  - and the microscopic states, intended as the values of the position and speed of all the
atoms of the gas. An equilibrium state is a macroscopic state in which
the value of the thermodynamical quantities are constant in time. This
does not mean that the system is frozen: the atoms of gas do not stop
moving, so that the \ microscopic state is changing. There is no contradiction, because to one macroscopic state correspond many
microscopic states. For instance by reversing the speed of each
of the atoms of gas, the values of the thermodynamical quantities is not not
modified: one changes the microscopic state but not the macroscopic one.

With this definition of equilibrium state in minds, it is  im\-portant to stress a tautology: equilibrium
has sense only because there exists an a priori notion of time, according to which one checks that the measured values of the
thermodynamical quantities are indeed constant. Time defines equilibrium. The thermal time hypothesis consists in reverting the
proposition: from the notion of equilibrium, one extracts time.  More exactly, starting from a physical system in a given state, one builds a
time-flow such that the state one has started with is precisely an equilibrium state with respect to the equilibrium notion defined by
this flow. There are two obvious difficulties:
\begin{enumerate}
\item  first one needs to characterize the would-be equilibrium states, that
is those states among all the states of a system for which there
exists a time-flow making them equilibrium states;
\item second, for the thermal time hypothesis to be of any use, one should be
able to extract \textit{explicitly }the time-flow from the knowledge
of a would-be equilibrium state.
\end{enumerate}

The characterization of states that could be equilibrium states exists, and is mathematically formulated in terms of
the so called \emph{KMS conditions.} These are mathematical relations involving the state under
consideration, the algebra of observables of the system and a group of
automorphisms $\alpha_t$ (see
section II.2 for the definition of group of automorphisms).  If the KMS
conditions are satisfied, then the state has, with respect to ${\alpha}_t$ the same properties as an equilibrium state
of a physical system whose quantum time-evolution would be given by the
time-as-abstract-flow $\alpha_t$.

Mathematics also provide a way to solve the second difficulty: the
modular theory of Connes, Tomita, Takesaki makes possible to extract from the knowledge of an algebra
of observables and a state ${\Omega}$ (both satisfying technical requirements) a 1-parameter group of
automorphisms $\alpha_t^\Omega$, called \textit{modular group}, such
  that the state $\Omega$ is KMS with respect to $\alpha_t^\Omega$.
This is not an easy piece of mathematics, so we shall not try to
explain it here. Let us simply stress that the modular theory mainly
deals with two simple but fundamental properties of the physical observables:
\begin{itemize}
\item  one is the algebraic structure of observables, namely the fact that
observables can be added and multiplied, this multiplication being
noncommutative (an algebra of quantum observables is a noncommutative
algebra: position $X$ multiplied by momentum $P$ is not the same as $P$
multiplied by $X$);
\item  the other is the fact that measured quantities are real numbers, opposed
to complex numbers. This is not a trivial requirement, since in quantum mechanics complex numbers play an important role, but
they do not come as the possible value of the measurement of a physical quantity (technically speaking: a quantum observable is a selfadjoint
operator). The mathematical translation of this requirement is that
the algebra of observables comes equipped with a tool allowing to separate
real quantities from complex ones, called an involution (for instance the involution on the set of complex
numbers is the usual complex conjugation: a real number is fully
characterized as a complex number which is equal to its conjugate).
\end{itemize}

To summarize, assuming one has at his disposal a notion of equilibrium state, the modular theory allows to extract from the algebra of
observables of a system a group of automorphisms that has all the properties one could expect from a bona fide time-as-abstract-flow. Furthermore,
this flow is unique: a state is a state of equilibrium only with respect to one modular flow. We shall call this
time-as-abstract-flow the \textit{thermodynamical time, } or \textit{thermal time }for short. To fully realize the program announced
at the beginning of this section, it remains to turn the abstract flow
of thermal time into a geometrical flow. This is the question we
examine in the next
section.

\section{Figures of thermal time}

For clarity let us repeat how the thermal time hypothesis proposes to solve the problem of the uniqueness of the classical limit of time-flow
discussed above. If one knows the algebra of observables of quantum gravity, the thermal time hypothesis allows to associate to an
equilibrium quantum state of the gravitational field a \textit{unique} thermodynamical time. Said differently, thanks to thermodynamics, one
is able to reduce the multiplicity of choice of time-flows at the quantum level (recall that a quantum state of the gravitational field
is a superposition of many possible choices of time-flow) to one single thermal-time flow. The bet then consists in claiming that this
thermal -time flow coincides with a geometrical flow of proper time. The claim immediately raises (at least !) three questions:

\begin{enumerate}
\item  what is the algebra of observables of quantum gravity ?
\item  the thermal-time flow is an abstract flow of automorphisms. How can
one retrieve from it a geometrical flow of proper time ?
\item in what sense is the thermal time unique ? What happens if one starts
with another equilibrium state ?
\end{enumerate}

\subsection{ From an abstract flow to a geometrical flow}

Let us examine the first points. For the moment, none of the tentative
theories of quantum gravity (e.g. loop quantum gravity, string theory)
clearly proposes a definition of the algebra of observables. Thus one
cannot test the pertinence of the thermal time hypothesis in this context. However one can check that, in physical situations where the
modular group (discussed in section III.2) is known, then its interpretation as a thermal-time makes sense.

The most famous example where this actually
happens is the so called Unruh effect. The later states that in
Minkowski space-time, the quantum vacuum is not seen as void by a uniformly accelerated
observer. It is seen has an equilibrium
state with temperature T proportional to the acceleration. This means the following: assume
Minkowski space-time is filed with a quantum field,
for instance an electro\-magnetic
field.{\footnote{To avoid confusion: we consider a quantum field theory, like
electromagnetism, defined on a background flat space-time. We do not
consider the gravitational field, that would curve the flat
space-time.}}
For a static observer, the field is in its vacuum state, meaning the observer does not see any photons (we do not
take into account the fluctuations of the vacuum). For an accelerated
observer on the contrary, the field is an equilibrium state, that is the observer sees photons, whose energy is similar as if
the observer was at rest, but immersed in a thermal bath of photons at temperature $T$.

Let us now interpret the Unruh effect at the light of the thermal time hypothesis: because no signal can go faster than light, a uniformly accelerated observer may exchange information,
that is send a signal and receive an answer, only with a ``receptor''  located into the region enclosed
by the two light-lines asymptotic to its trajectory (see figure
2). This region is called \textit{a Rindler wedge}, and is
traditionally denoted $W$. This region is said \textit{causally
  connected} to the uniformly accelerated ob\-server.
One then considers the set of observables localized into $W$. Their
precise definition is given in algebraic
quantum field theory. For our purposes, it is
enough to know that such localized observables form
an algebra, denoted $\A(W)$, and
represent the amount of observable
information accessible to the uniformly accelerated observer. The
modular group defined by $\A(W)$ and the vacuum state
$\Omega$ of the quantum field theory has
been computed in \cite{Bisognano:1976kl}. By definition, this is a
group of automorphisms of $\A(W)$, but it turns out that this group also has a geometrical
action.{\footnote{To understand how an automorphism can have a geometrical action, one needs
to go deeper into the definition of the algebra of local observables $\A(W)$. The latter carries a representation
of the Poincar\'e group, so that a Poincar\'e transformation can be
viewed both as a transformation on space-time and as an autormorphism
of $\A(W)$. However not all the automorphisms
of $\A(W)$ come from a Poincar\'e
transformation. The remarkable point here is that the modular group of
automorphisms precisely consists in automorphisms coming from the
action of the Poincar\'e group.}
And this action precisely coincides with the
flow of proper time of an accelerated observer. Therefore the abstract
flow of thermal-time coincides with the geometrical flow of proper
time of a uniformly accelerated observer.

A similar analysis can be done for other regions of Minkowski space-time, like double-cones (figure 3). These are regions of
particular interest since in the analysis of the Unruh effect above,
we kept silent a important point: the Rindler wedge is the region causally
connected to a uniformly accelerated observer whose \textit{lifetime is
infinite.} Indeed, the line of universe depicted in figure 2 extends
from \ $-$ ${\infty}$ (where it is asymptotic to a light ray
propagating towards negative \textit{X}) \ to + ${\infty}$ (idem, with
light propagating towards positive \textit{X}). The region causally
connected to a uniformly accelerated observer with finite lifetime
is no longer a Rindler wedge but the double-cone obtained as the
intersection of the future cone of the birth of the observer, with the
past cone of its death (see figure \ref{diam}). As in the wedge case, one finds that the
modular flow of the algebra of observables localized inside the
double-cone has a geometrical action, which coincides with the
trajectory of a uniformly accelerated observer being born and dying at
the tips of the cone.}

\begin{figure}[h*]
\begin{center}
\mbox{\rotatebox{0}{\scalebox{.6}{\hspace{1truecm}\includegraphics{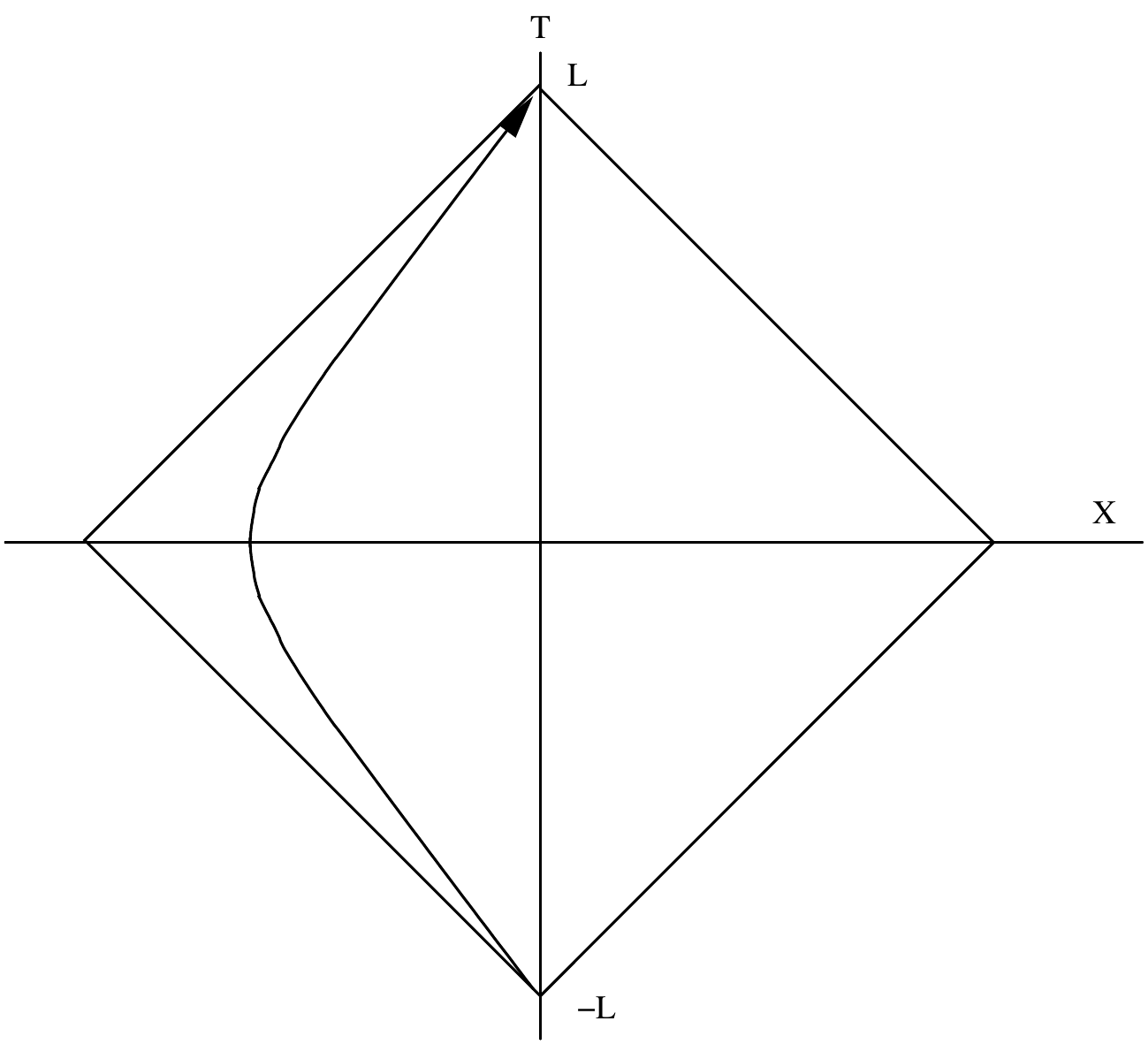}}}}
\begin{minipage}{.85\linewidth}{\caption{\label{diam}\emph{Double-cone}: the region of Minkowski space-time causally connected to a
uniformly accelerated observer with finite lifetime is a double -cone,whose size L is related to the lifetime of the observer.}}\end{minipage}
\end{center}
\end{figure}

 \vspace{-.0truecm}} The difference with the wedge case is that the temperature not only
depends on the acceleration, but also on the lifetime of the observer
\cite{Martinetti:2003sp, Martinetti:2009ff}.
The consequences of this dependence on causality has been investigated
in \cite{Martinetti:2007kb, Martinetti:2007fu}: since the
temperature depends on the lifetime, can an observer deduce the
instant of its death from a measurement of the temperature ? The answer
is no: causality is protected by the fourth Heisenberg uncertainty
relation $\Delta E \Delta t \geq \hbar$, where $E$ is the energy, $t$
the time and $\hbar$ the Planck constant. Indeed, the difference in temperature between the
eternal and non-eternal cases is smaller than the smallest
temperature $\frac{\hbar}{k\Delta t}$ that an observer with lifetime
$\Delta t$ can measure ($k$ is the Boltzmann constant). In other
term, a non-eternal observer does not live long enough to realize he
is not eternal.

 Both the examples of the wedge and the double-cone illustrate how
 the thermal time hypothesis can reconcile the time-as-abstract flow of
quantum mechanics with the time-as-geometrical -flow of relativity
discussed in section II. The abstract flow of automorphisms and the
geometrical flow in space-time are two ways of flowing that are not
contradictory. They are two figures of the same ``time'', like the proper times of different
observers are various figures of a same ``geometrical
time''. However one should not believe that this situation
is generic.

\subsection{Is time necessarily flowing ?}

Nothing in the formalism of the modular group guarantees that the
flow of thermal time should admit a geometrical representation. Even in this
case, nothing guarantees that this flow can be identified with a flow of proper time (for instance a space-like flow would correspond to an
observer moving faster than light, hence non-physical).

Recently an example has been worked out in which the flow of thermal
time is not purely geometric \cite{Longo:2009uq, Casini:2009zm}. One considers again a double-cone
region in a two-dimensional Minkowski space-time, but the quantum field theory
is assumed to satisfy particular symmetry conditions (conformal
invariance {\footnote{In mathematics, a conformal map is a function which preserves angles.}}
and specific boundary conditions). As a consequence, any observable localized into
the double-cone is completely determined by its value on two intervals
on the axis (see figure 4).

\begin{figure}[h*]
\begin{center}
\vspace{-4truecm}
\mbox{\rotatebox{0}{\scalebox{.6}{\hspace{1truecm}\includegraphics{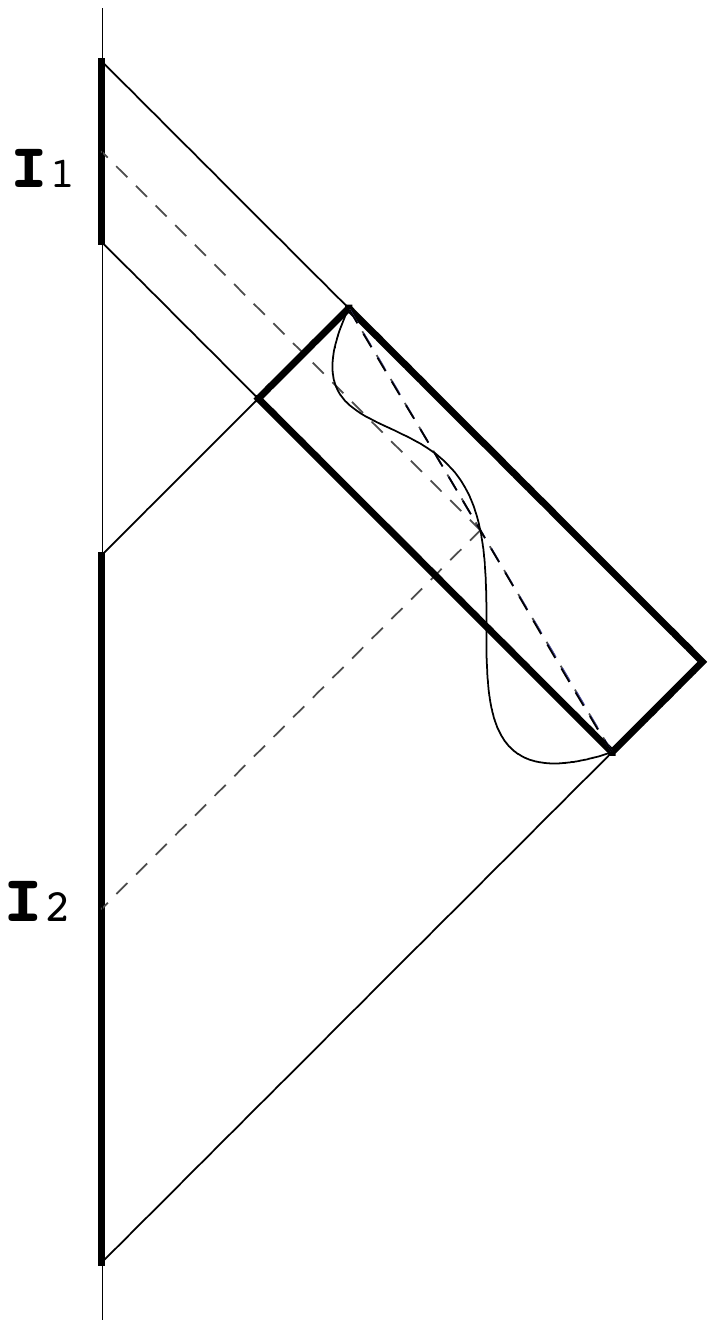}}}}
\begin{minipage}{.85\linewidth}{\vspace{-4truecm}
\caption{\label{diam} \emph{Double-cone in a bi-dimensional boundary conformal field theory}: the value of the field inside the
double-cone is completely determined by its value on the two intervals $I_1, I_2$. The curve represents
the line of universe associated to the thermal time generated by a
certain state (Longo ad -hoc state) of the algebra of observables of
the double-cone. This thermal-time is the proper time of an
observer with non-constant acceleration. This can be seen by noticing
that at various points along the curve, the tangent vector is vertical,
meaning that the speed changes signs several times (compare to the
double-cone in Minkowski space-time of figure 2). Notice that the line of universe is in reality
contained within the double-cone: in the figure,  the oscillation of the curve around the diagonal of the double-cone (dashed line) has
been artificially exaggerated, so that to make the effect visible on the
plot.}}
\end{minipage}
\end{center}
\end{figure}

There exists a certain ad-hoc quantum state
(not the vacuum however), whose associated thermal time has a
representation in terms of a geometrical flow of proper
time. Notice that this flow is not the flow of proper time of a uniformly accelerated observer,
but of an observer whose acceleration is not constant, and even change sign (see fig. 4).

 Let us consider now the thermal time defined by the vacuum state. One finds that the modular flow of the vacuum, as flow of automorphisms,
combine the above geometric action with another non-geometric action, namely an automorphism whose action amounts to mix the fields
in the two intervals  $I_1, I_2$. As explained in the footnote e p.8,
an automorphism - being an object living at the level of the algebra of
observables of a system - does not need to admit a geometrical
representation. The mixing term of the modular group of the vacuum
state is precisely one of these ``purely algebraic'' automorphisms. The
thermal time is thus flowing as an abstract flow, but its concrete
realization is no longer simply a geometrical-flow: time is
simultaneously ``something that flows geometrically'' and
``something that mixes the components'' in
a non -geometrical way.

This is a good place to address point 3 in the list of questions made at the beginning of this section. To what extent are these two thermal
times  - the purely geometrical one defined by the ad-hoc state, and the partially geometric one defined by the vacuum  - two figures of a
still unique ``abstract time'' ? In a seminal result of operator algebra \cite{Connes:1973uq}, Connes has shown that the
modular flows defined by different states are equal up to inner automorphism. Without entering the details, this means that the two
thermal times are not ``so much different'', they
only differ by a unitary transformation. In other terms, given an
algebra of observables, up to these inner automorphisms, it makes sense
to talk about a\textit{unique thermal time.}

\section{Conclusion: is there more light at noon or midnight ?}

To conclude, let us illustrate these various figures of time, not always
easy to deal with, coming back to the analogy with the usual time on
Earth. The Unruh effect described above can be interpreted as follows: to
the question
\begin{center}
Why is there something rather than nothing ?
\end{center}
the accelerated observer (who sees the quantum vacuum as populated with quanta at temperature $T$, e.g. photons) will answer to his inertial
partner (who sees the vacuum as void): ``there is something for me because I am accelerating, and there is nothing for
you because you are inertial''. It may sound strange that
the answer to the question
\begin{center} Is there anything rather than nothing ?
  \end{center}
depends on the acceleration and thus, since an acceleration characterizes a proper time, depends on the temporal
framework in which the question is asked. In terms of time-zones and time-systems, this phenomenon has in fact
nothing mysterious: to the question
\begin{center} 
Is there more light at day or night ?
  \end{center}
an observer whose clock is regulated on the solar-time will answer ``noon''. Indeed, in Kathmandu like in Haiti,
there is more light at day than at night. On the other hand, an
observer regulated on the lunar-time will answer: ``it depends''. His lunar-day (that is the moment of time
when the Moon is visible in its sky) may correspond to a solar day
(when the Moon is visible by day), but the lunar-day may also
corresponds to a solar-night (when the Moon is visible at night). In this case, there is more
light at lunar-night than at lunar-day.

Quantizing gravity means loosing the possibility to associate to an
observer a unique time-system. Each observer may happen to be
regulated on whatever time-system (solar, lunar, jupiterian etc). To
the question ``when is there more light
?'', the same observer may answer ``at day'', and one understands that his quantum superposition
of clocks has just collapsed to a clock regulated on solar-time, or
he may answer ``I do not know, your question has no
meaning'', and one understands that his quantum clock has
collapsed to a clock regulated on e.g. the moon-time.

However, if one forgets the quantum aspect of gravity, this observer
lives according to his unique proper time, which is regulated on a a precise time-system. The question of the unicity of
the classical limit thus consists in reconciling the
superposition of all possible regulations - at the quantum level
 - with the uniqueness of the local time-system. 

Connes and Rovelli propose that time-flow is an emergent notion,
stemming from the algebraic structure of the observables, together with
a state of the physical system under consideration. Rather than
determining the state of the physical system using the point of view of
one  precise observer, attached to an a priori given
time-system (You, whose clock is solar, do you see
more light at night or day ?), one considers that the
state of the system induces a particular time-flow: the state
``day has more light than night''  implies
solar time; the state ``sometimes days has more light than
night, sometimes no'' implies lunar time.

\small{\bibliographystyle{abbrv}
\bibliography{/Users/pierremartinetti/physique/articles/Bibdesk/biblio}}
\end{document}

%% file: preambulesobre.tex
\usepackage[english]{babel}
\usepackage[isolatin]{inputenc} 

\usepackage{amsfonts,amssymb,amsmath,bbm}
\usepackage{fontenc,indentfirst, delarray}
\usepackage{graphicx,graphics,xcolor,epstopdf,epsf}
\DeclareGraphicsExtensions{eps,ps,p.gz,pdf}
\usepackage{pdfsync}
\usepackage[pdfstartview=FitH]{hyperref}
\usepackage{setspace}

\usepackage{footmisc}
\usepackage{fancyheadings}
\renewcommand{\thesubsection}{{\Roman{section}.\arabic{subsection}}}
\renewcommand{\thesection}{{\Roman{section}}}
\makeatletter
\renewcommand{\section}{\@startsection {section}{1}{\z@}%
             {-3.5ex \@plus -1ex \@minus -.2ex}%
             {2.3ex \@plus.2ex}%
             {\normalfont\normalsize\sffamily\bfseries}}
\renewcommand{\subsection}{\@startsection {subsection}{1}{\z@}%
             {-3.5ex \@plus -1ex \@minus -.2ex}%
             {2.3ex \@plus.2ex}%
             {\normalfont\normalsize\sffamily\emph}}

%% file: commandes.tex
\definecolor{bleuvert}{rgb}{.1,.5,.4}
\definecolor{light-gray}{gray}{0.95}
\definecolor{gray}{gray}{0.75}
\definecolor{violet}{rgb}{0.4,0.,0.3}
\definecolor{jaune}{rgb}{0.8,0.6,0.1}
\definecolor{cvert}{rgb}{0.8,0.6,0.5}

 \definecolor{couleur}{rgb}{.1,.5,.4}
  
\def\begf{\begin{frame}}
\def\enf{\end{frame}}

\def\begz{\begin{itemize}}

\def\endz{\end{itemize}}

\def\lp{\left(} 
\def\rp{\right)} 
\def\dm{\lp\begin{array}}	
\def\fm{\end{array}\rp}
\def\m2{M_2 \lp \cc \rp}
\def\m3{M_3 \lp \cc \rp}

\def\cc{{\mathbb{C}}}

\def\A{{\mathcal A}}

\def\xo0{\omega^0_x}
\def\yo0{\omega^0_y}

\def\xo0{x_\omega^0}
\def\yo0{y_\omega^0}

\def\fm{\Phi(x^\mu)}

\def\dm{\partial_\mu}

\def\dmm{\left(\begin{array}}
\def\fmm{\end{array}\right)}